\documentclass[12pt,epsf]{article}

\usepackage[dvipdfmx]{graphicx,color}
\usepackage{amsmath}

\usepackage{graphicx}
\newlength{\subfigwidth}
\newlength{\subfigcolsep}
\setkeys{Gin}{width=\subfigwidth}
\makeatletter 
 
\@addtoreset{figure}{section} 
\makeatother 
\usepackage{subfigure}

\newcommand{\beq}{\begin{equation}}
\newcommand{\eeq}{\end{equation}}
\newcommand{\bea}{\begin{eqnarray}}
\newcommand{\eea}{\end{eqnarray}}
\usepackage{bm}

\begin{document}
\thispagestyle{empty}
\vspace*{-15mm}
{\bf OCHA-PP-343}\\

\vspace{15mm}
\begin{center}
{\Large\bf
A simulation of hydrodynamics \\on non-commutative space \\
}
\vspace{7mm}

\baselineskip 18pt
{\bf Tetuya Kawamura$^{1}$, Anna Kuwana$^{2}$, Yusaku Nagata$^{1}$,\\
Mayumi Saitou$^{1}$ and Akio Sugamoto$^{1, 3}$}
\vspace{2mm}

{\it
$^1$Faculty of Sciences, \\
   Ochanomizu University, Tokyo 112-8610, Japan}\\
{\it
$^2$Information Technology Center, \\
Ochanomizu University, Tokyo 112-8610, Japan}\\
{\it
$^3$Tokyo Bunkyo Study Center, \\
The Open University of Japan, Tokyo 112-0012, Japan}
\vspace{10mm}
\end{center}
\begin{center}
\begin{minipage}{14cm}
\baselineskip 16pt
\noindent
\begin{abstract}
A simulation of the hydrodynamics on the two dimensional non-commutative space is performed, in which the space coordinates $(x, y)$ are non-commutative, satisfying the commutation relation $[x, y]=i \theta$.  The Navier-Stokes equation has an extra force term which reflects the non-commutativity of the space, being proportional to $\theta^2$. 
This parameter $\theta$ is related to the minimum size of fluid particles which is implied by the uncertainty principle, $\Delta x \Delta y \ge \theta/2$. To see the effect of this parameter on the flow, following situation is considered. 
An obstacle placed in the middle of the stream, separates the flow into small slit and large slit, but the flow is joined afterwards in the down stream.  For the Reynolds number 700, the behavior of the flows with and without $\theta$ is observed to differ, and the difference is seen to be correlated to the difference of the activity of vortices in the down stream.  The oscillation of the flow rate at the small slit diminishes after the certain time in the usual flow when the ``two attached eddies" appear. In the non-commutative flow this two attached eddies appear from the beginning and the behavior of the flows does not fluctuate largely.  The irregularity in the  flow existing in the beginning disappears after the certain time.
 \end{abstract}
\newpage
\end{minipage}
\end{center}
\baselineskip 18pt
\def\thefootnote{\fnsymbol{footnote}}
\setcounter{footnote}{0}
\newpage
\section{Hydrodynamics on non-commutative space}
In the classical mechanics, the phase space volume occupied by dynamical systems is preserved in time.  This is the statement of the Liouville theorem, shown by using the Hamilton dynamics:
\begin{eqnarray}
\dot{x}=\frac{\partial H}{\partial p}=\{x, H\}, ~~\dot{p}=-\frac{\partial H}{\partial x}=\{p, H\}.
\end{eqnarray}
If we replace the phase space $(x, p)$ by the spacial coordinates $(x, y)$, then the volume of the real space occupied by particles is preserved in time.  This gives the incompressible hydrodynamics:
\begin{eqnarray}
v_x=\dot{x}=\frac{\partial \varphi }{\partial y}=\{x, \varphi\}, ~~v_y=\dot{y}=-\frac{\partial \varphi}{\partial x}=\{y, \varphi\}, 
\label{2d velocity}
\end{eqnarray}
where $\{A, B\}$ is the Poisson bracket, and $\varphi(x, y)$ is the stream function of the incompressible fluid which guarantees the continuity of the flow, $\nabla \cdot \bm{v}=0$. 
If the flow is incompressible perfect fluid, the $\varphi(x,y)$ is regarded as the stream function.
 Here we assume the density of the fluid $\rho$ is constant in time.  It is easily understood that if we consider the three dimensional fluid, Eq. (\ref{2d velocity}) is replaced by
\begin{eqnarray}
v_i=\dot{x}_i=\{ x_i, \varphi_1, \varphi_2\}, ~(i=1\sim3), \label{3d velocity}
\end{eqnarray}
where the right hand side $\{A, B, C\}$ is the Jacobian $\partial(A, B, C)/\partial(x_1, x_2, x_3)$, being called Nambu bracket at present \cite{Nambu dynamics} . 
The advective term which describes the temporal change of a physical quantity $O$ by floating with the flow, is represented in terms of Poisson or Nambu bracket as  
\begin{eqnarray}
(\bm{v} \cdot \nabla)\ O=\{O, \varphi_1, \cdots \}.
\end{eqnarray}
Therefore, the fluid dynamics can be given as 
\begin{eqnarray}
\frac{D \bm{v}}{Dt} \equiv \frac{\partial \bm{v}}{\partial t}+\{\bm{v}, \varphi_1, \cdots \}=\frac{1}{\rho}\bm{F},
\end{eqnarray}
where $\bm{F}$ is the force acting on the fluid.
Now the Navier-Stokes equation can be described in terms of Nambu-Poisson brackets:
\begin{eqnarray}
&&\rho\left(\{x_i, \dot{\varphi}\}+\big\{\{x_i,\varphi\},\varphi\big\}\right)
+\epsilon^{ij}\{p,x_j\}-\eta\Delta\{x_i,\varphi\}=0\ (\mathrm{for\ 2\ dimensions}),\quad\\
&&\rho\left(\{x_i, \dot{\varphi_1},\varphi_2\}+\{x_i, \varphi_1, \dot{\varphi_2}\}
+\big\{\{x_i,\varphi_1, \varphi_2 \},\varphi_1,\varphi_2 \big\}\right)\nonumber\\
&&\quad\quad\quad\quad\quad\quad
+\epsilon^{ijk}\{p,x_j,x_k\}-\eta\Delta\{x_i,\varphi_1,\varphi_2\}=0
\quad(\mathrm{for\ 3\ dimensions}), 
\end{eqnarray}
where $p(x, y, t)$ is the pressure and $\eta$ is the viscosity. 
This was done by Nambu in his last work \cite{Nambu 1}.

Now we introduce the non-commutativity in space coordinates, and find the additional terms which will appear.  This was carried out in \cite{Saitou 1} where the non-commutativity is introduced by replacing the Nambu-Poisson brackets with the Moyal brackets \cite{Moyal}.  We discuss the two dimensional hydrodynamics in this paper, so the Moyal bracket becomes
 \begin{eqnarray}
[A, B]_P \xrightarrow{replacement} \frac{1}{i\theta}[A, B]_M,
\end{eqnarray}
where
\begin{eqnarray}
[A,B]_M&\equiv&A*B-B*A\\
A*B&\equiv&\exp\left( \frac{i\theta}{2}\epsilon_{ab}
\frac{\partial^2}{\partial y^a \partial z^b}\right) A(y)B(z)\bigg|_{y,z\rightarrow x}
\end{eqnarray}
This replacement is equivalent to consider the space coordinates $(x, y)$ to be the operators $(\hat{x}, \hat{y})$ satisfying the commutation relation $[\hat{x}, \hat{y}]=i \theta$, the $\theta$ is a constant parameter. From this, the uncertainty relation is derived:
\begin{eqnarray}
\Delta x \Delta y \equiv \sqrt{\langle (\Delta x)^2 \rangle} \sqrt{\langle (\Delta y)^2 \rangle}\ge \frac{1}{2}\theta.
\end{eqnarray}
The three dimensional case is a little more complicated.  See the details in \cite{Saitou 1}.

After the replacement of the Poisson brackets with the Moyal products, new terms arise only from the advective term, that is
\begin{eqnarray}
&&\frac{1}{i\theta} [v_i,\varphi]_M=
\{v_i,\varphi\}_P\nonumber\\
&&-\frac{\theta^2}{24}
\left( \frac{\partial}{\partial y^a}\frac{\partial}{\partial z^b}
-\frac{\partial}{\partial y^b}\frac{\partial}{\partial z^a} \right)^3
v_i(y^a,y^b) \varphi(z^a,z^b)|_{y,z \rightarrow x}
+O(\theta^4). 
\label{1-9}
\end{eqnarray}

Now, the Navier-Stokes equation of the hydrodynamics on the 2 dimensional non-commutative space is given by \cite{Saitou 1}
\begin{eqnarray}
&&\rho\frac{D \bm{v}}{Dt}+\nabla p -\eta \Delta \bm{v}=\bm{K}\\
&&K_i=\rho\ \frac{\theta^2}{24}
\left( \frac{\partial^3 v_i }{\partial x^3 }\frac{\partial^2 v_x  }{\partial y^2}
+3\frac{\partial^3 v_i}{\partial^2 x \partial y} \frac{\partial^2 v_y }{\partial y^2}
+3\frac{\partial^3 v_i}{\partial x\partial y^2}\frac{\partial^2 v_x }{\partial x^2}
+\frac{\partial^3 v_i }{\partial y^3}\frac{\partial^2 v_y }{\partial x^2}
 \right)  \nonumber \\
 &&\quad\quad\quad+O(\theta^4).
\end{eqnarray}
The $\bm{K}$ is the new force term which reflects the non-commutativity of the space.

In the study of hydrodynamics, Reynolds number, Re $=\rho U_{\star}\ L_{\star} /\eta$ is important, which is the ratio of the kinetic term to the viscosity 
term in the equation of motion and it determines the gross behavior of the flow.  In the non-commutative hydrodynamics, another measure appears, which is the ratio of the extra force term to the kinetic term in the equation of motion, {\it i.e.} Sm $=(\theta/L_{\star})^2$.  Here, $U_{\star}$ and $L_{\star}$ is the typical velocity and length scale in the problem.

\section{Extra force $\bm{K} (\propto \theta^2)$ in the perfect fluid} 
In the perfect fluid without viscosity ($\eta=0$), it is easy to understand the behavior of the extra force $\bm{K}$ which is inherent to the non-commutativity.  

Let's examine the typical flows.

1) One dimensional flow; if $v_y=0$, and $v_x=f(x)$, then $\bm{K}=\bm{0}$. 

To consider the behavior of $\bm{K}$ near the sink, the source, the vortex, and the flow along the corner, the complex notations are useful.  Define $z=x+iy, \bar{z}=x-iy$, and their derivatives $\partial=\frac{1}{2}(\partial_x-i\partial_y)$, $\bar{\partial}=\frac{1}{2}(\partial_x+i\partial_y)$, and $v=v_x-iv_y$ and $K=K_x-iK_y$, then we have 
\begin{eqnarray}
K=-\rho \frac{2}{3} \theta^2 \partial \varphi \left(\overleftarrow{\partial} \overrightarrow{\bar{\partial}} - \overleftarrow{\bar{\partial}} \overrightarrow{\partial} \right)^3 \varphi(z, \bar{z}), 
\end{eqnarray}
where $\varphi(z, \bar{z})$ is the stream function.  Let's introduce the complex velocity potential $f(z)$ so that its real part is the usual velocity potential and its imaginary part is the stream function. Then, $\varphi(z, \bar{z})=\frac{1}{2i}(f(z)-\overline{f(z)})$, and $v=\partial f(z)$.

Now we have the following results:

2) Source or sink; $f(z)= m \ln z ~(m>0$ is a source, and $m<0$ is a sink.), and $K=2\rho~ \theta^2 m^2 \bar{z}/|z|^8$, or $\bm{K}=2\rho ~\theta^2 m^2{\color{black}\bm{r}}/r^8$.

3) Vortex; $f(z)= i\kappa \ln z ~(\kappa>0$ is clock-wise, and $\kappa<0$ is counter-clockwise), and $K=2\rho~ \theta^2 \kappa^2 \bar{z}/|z|^8$, or $\bm{K}=2\rho ~\theta^2 \kappa^2\ \bm{r}/r^8$.

4) Flow along the corner with angle $\alpha=\pi/n$;~$f(z)= c z^n$(c:~a real constant), and $\bm{K}=-\frac{1}{6} \rho~ \theta^2 c^2 \left( n(n-1)(n-2) \right)^2 (3-n)\ \bm{r}/r^{(8-2n)}$.

From 2), we understand that the additional force $\bm{K}$ disturbs the fluid to flow into the sink, but accelerates it to flow out from the source.  From 3), the force $\bm{K}$ is understood to expand the vortex in the radial direction.  

Therefore, the minimum size of the particle $\sqrt{\theta}$ implied by the non-commutativity gives the indication that the flowing into the small channel becomes difficult due to the finite size effect of the particle.  So, we may hope that the hydrodynamics on the non-commutative space gives the hydrodynamics of the granular materials.

In the above examples, the viscous force vanishes, $\eta \Delta \bm{v}=0$, and hence the behavior of $\bm{K}$ does not be altered, even if introducing the viscosity perturbatively as the expansion from the perfect fluid.

\begin{figure}[h]
  \setlength{\subfigwidth}{.25\linewidth}
  \addtolength{\subfigwidth}{-.25\subfigcolsep}
  \begin{minipage}[b]{\subfigwidth}
    \centering\subfigure[Flow field of source on the origin]{\includegraphics[width=3cm]{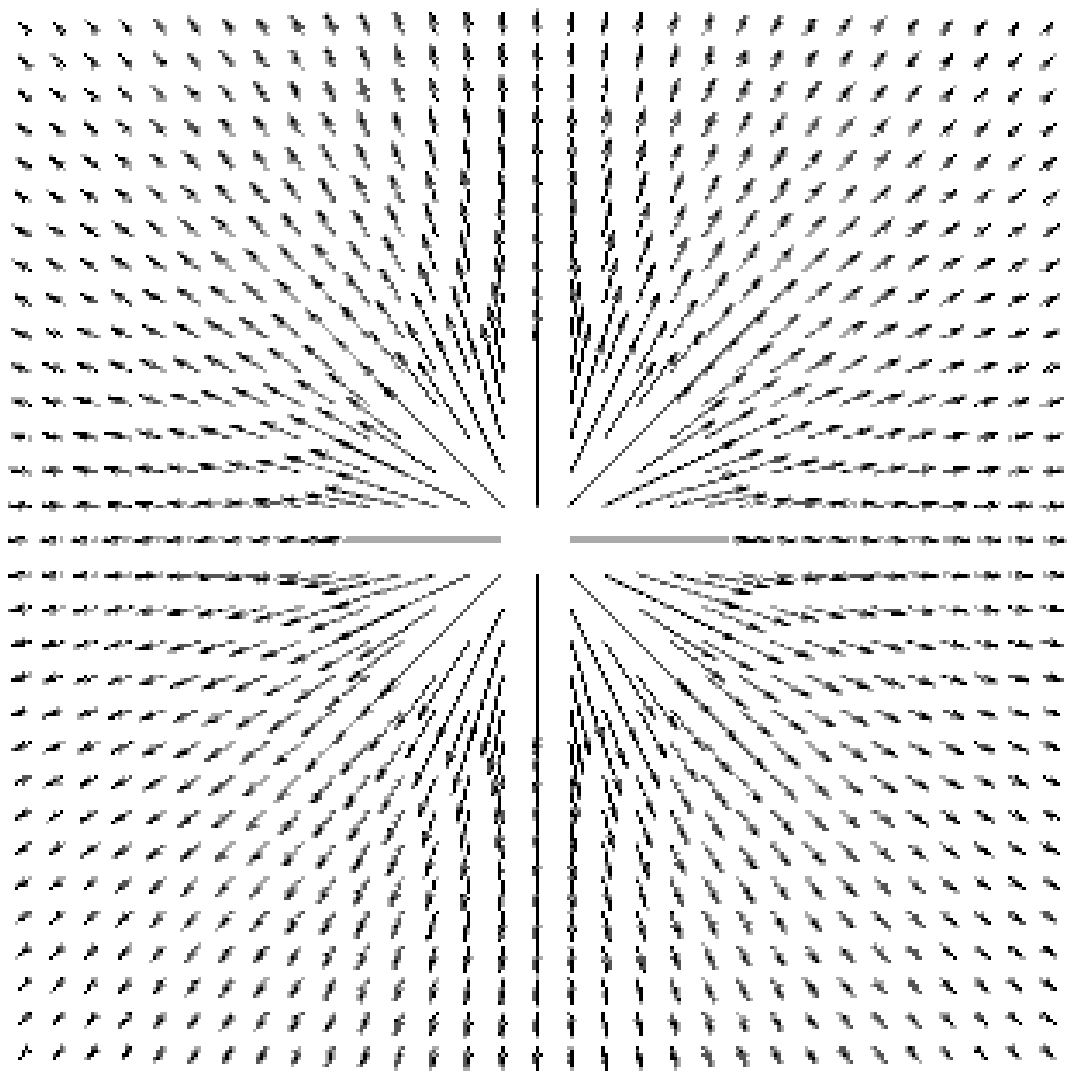}
\label{st1v}}
  \end{minipage}
  \begin{minipage}[b]{\subfigwidth}
    \setcounter{subfigure}{1}
    \centering\subfigure[Flow field of sink on the origin]{\includegraphics[width=3cm]{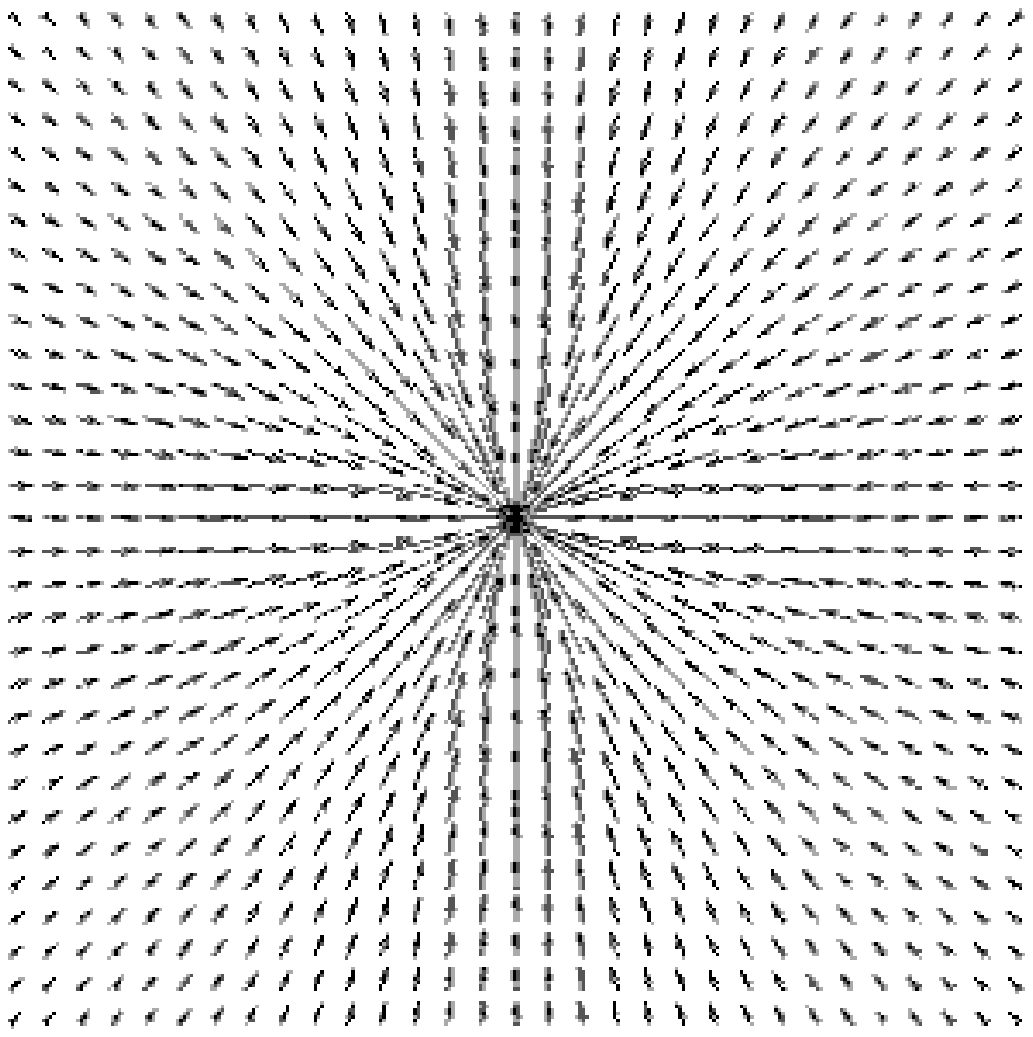}
    \label{st2v}}
  \end{minipage}
  \begin{minipage}[b]{\subfigwidth}
    \setcounter{subfigure}{2}
    \centering\subfigure[Flow field of vortex on the origin]{\includegraphics[width=3cm]{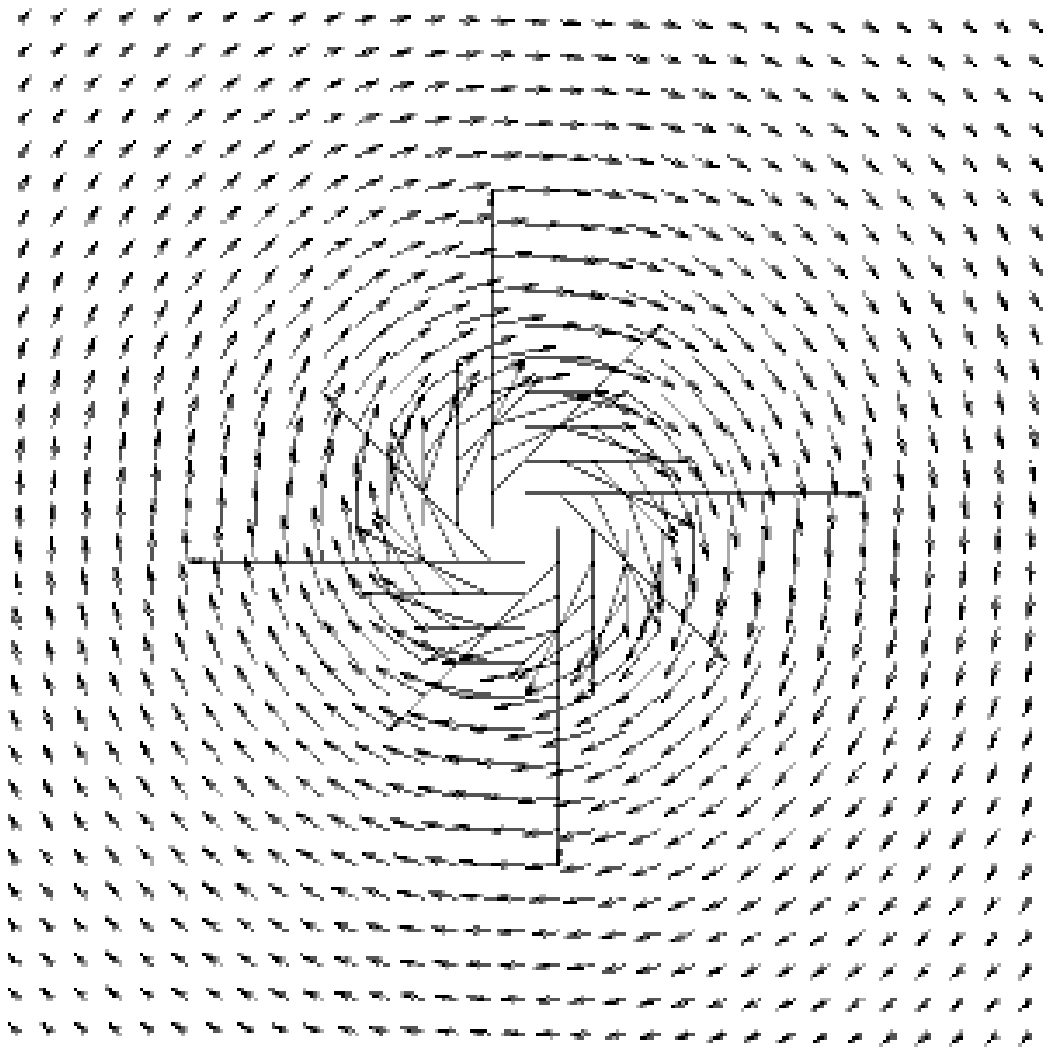}
    \label{st3v}}
  \end{minipage}
  \begin{minipage}[b]{\subfigwidth}
    \setcounter{subfigure}{3}
    \centering\subfigure[Flow of flow going around the board]{\includegraphics[width=3cm]{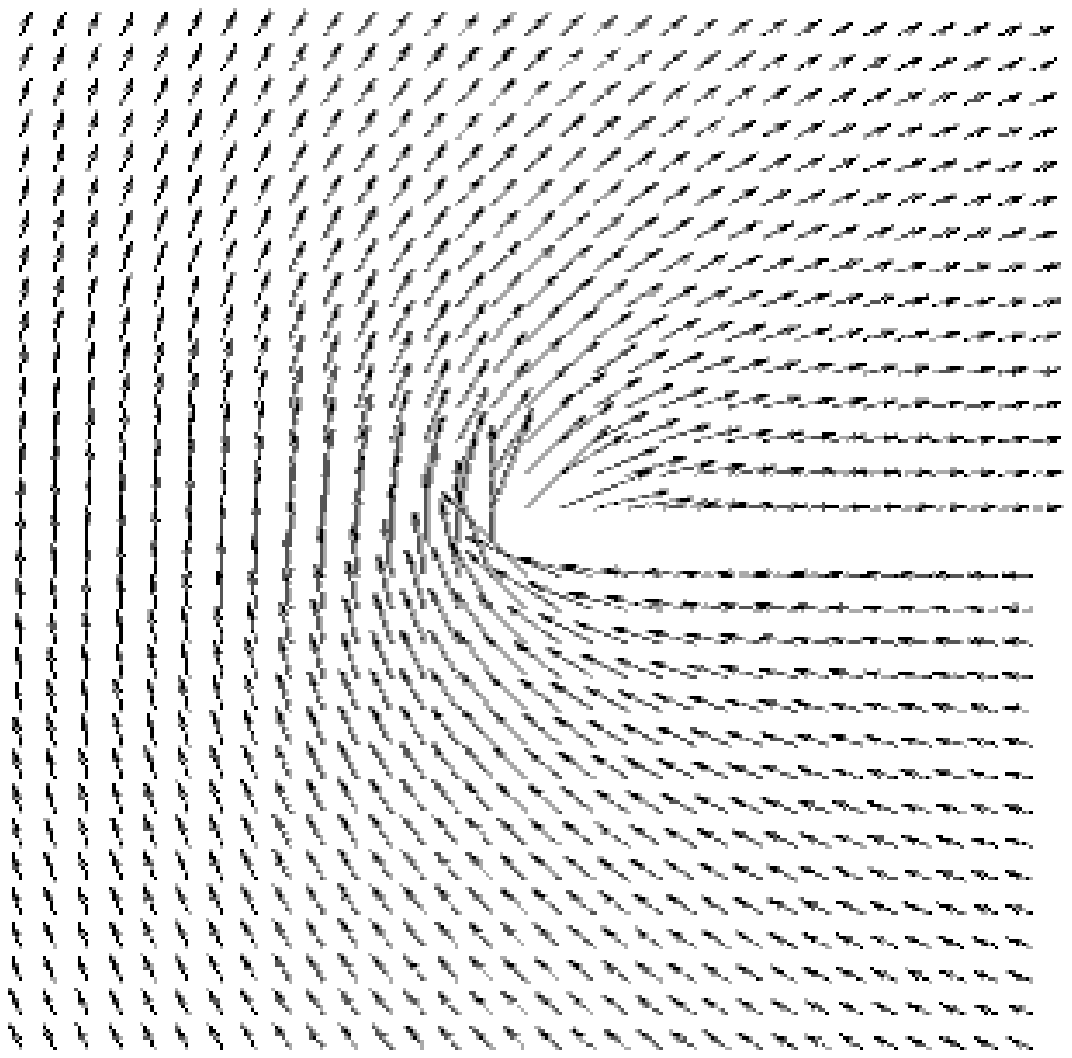}
    \label{st4v}}
  \end{minipage}
  \begin{minipage}[b]{\subfigwidth}
    \setcounter{subfigure}{4}
    \centering\subfigure[Force field of $\bm{K}$ induced by source of Fig.\ref{st1v}]{\includegraphics[width=3cm]{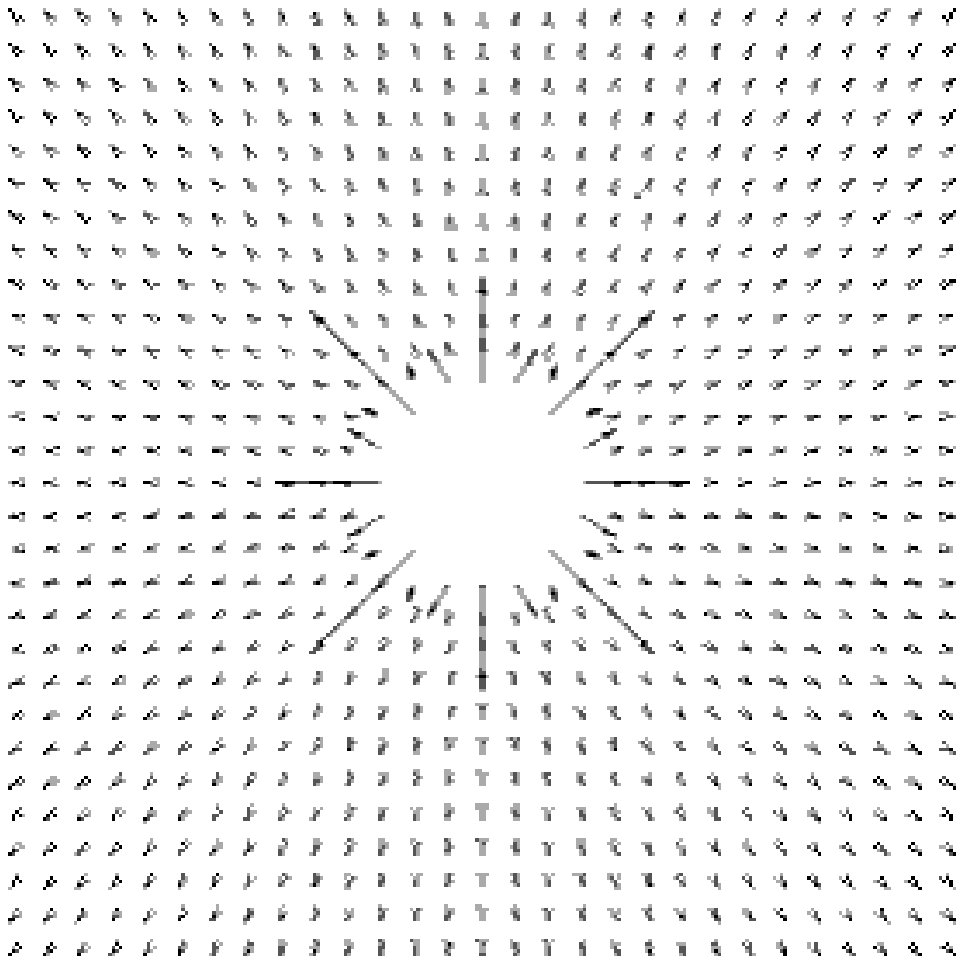}
    \label{st1k}}
  \end{minipage}
  \begin{minipage}[b]{\subfigwidth}
    \setcounter{subfigure}{5}
    \centering\subfigure[Force field of $\bm{K}$ induced by sink of Fig.\ref{st2v}]{\includegraphics[width=3cm]{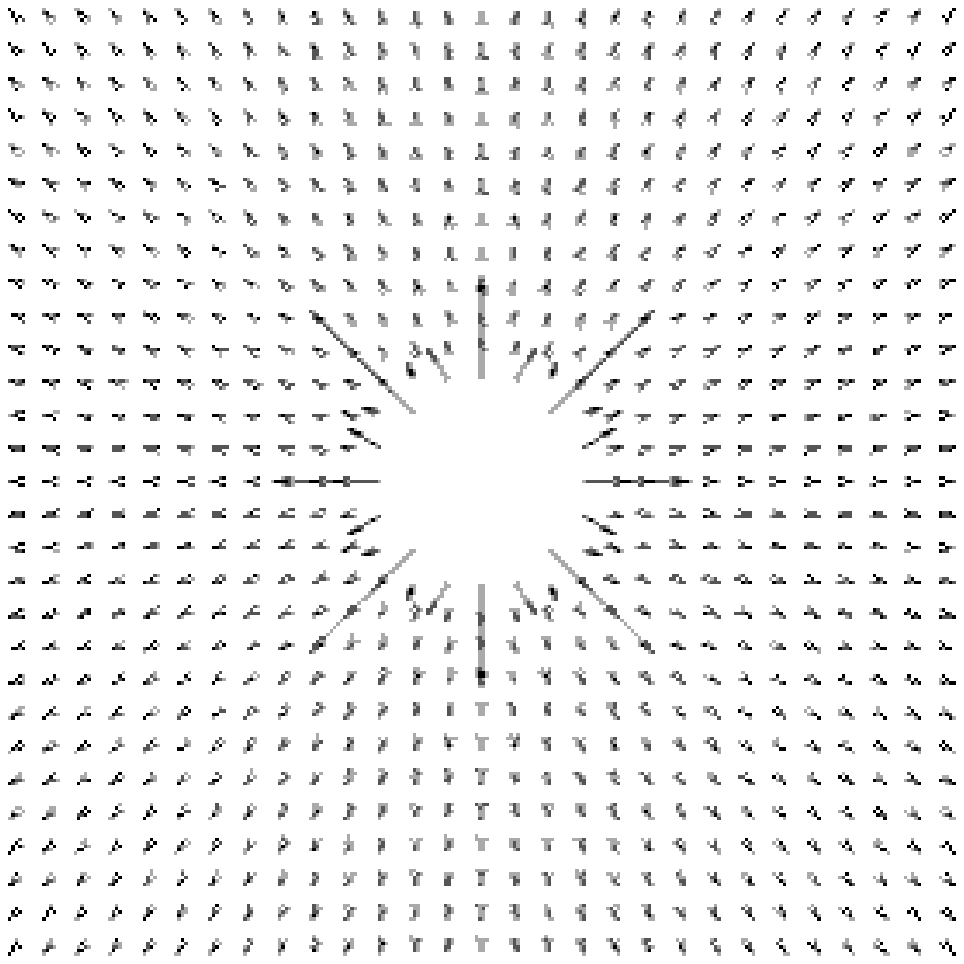}
    \label{st2k}}
  \end{minipage}
  \begin{minipage}[b]{\subfigwidth}
    \setcounter{subfigure}{6}
    \centering\subfigure[Force field of $\bm{K}$ induced by vortex of Fig.\ref{st3v}]{\includegraphics[width=3cm]{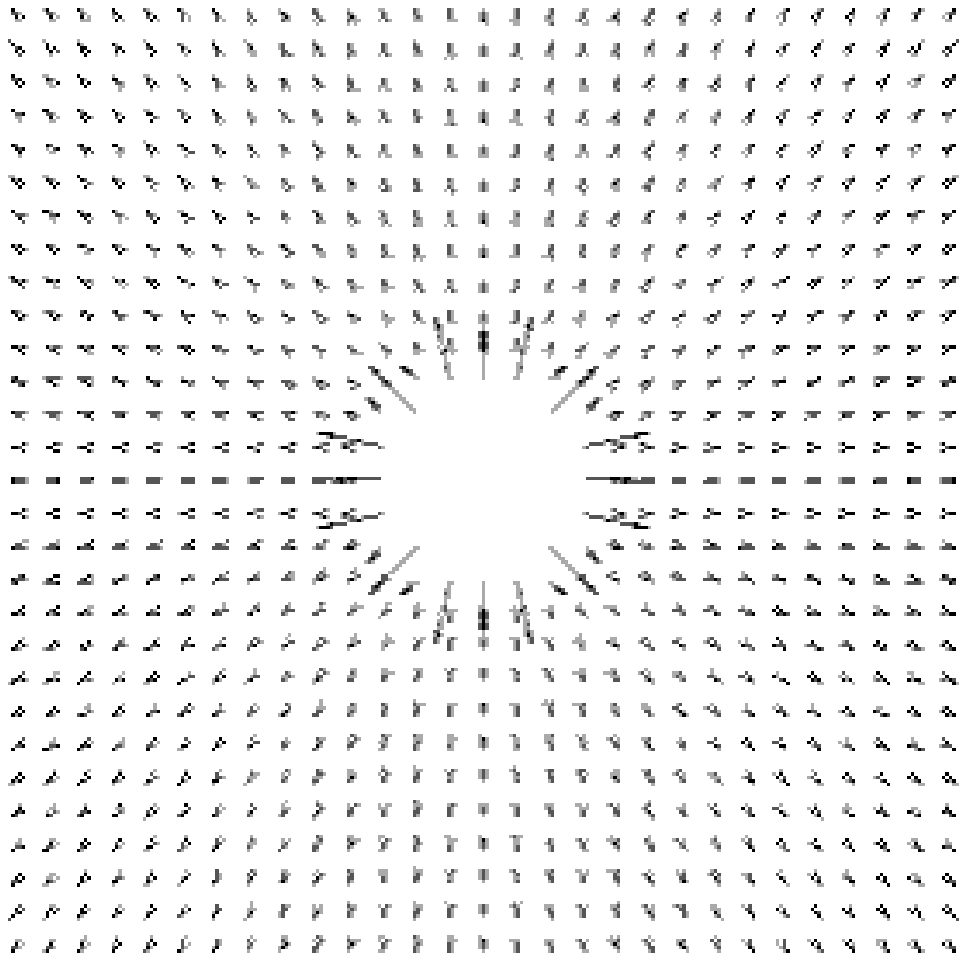}
    \label{st3k}}
  \end{minipage}
  \begin{minipage}[b]{\subfigwidth}
    \setcounter{subfigure}{7}
    \centering\subfigure[Force field of $\bm{K}$ induced by flow of Fig.\ref{st4v}]{\includegraphics[width=3cm]{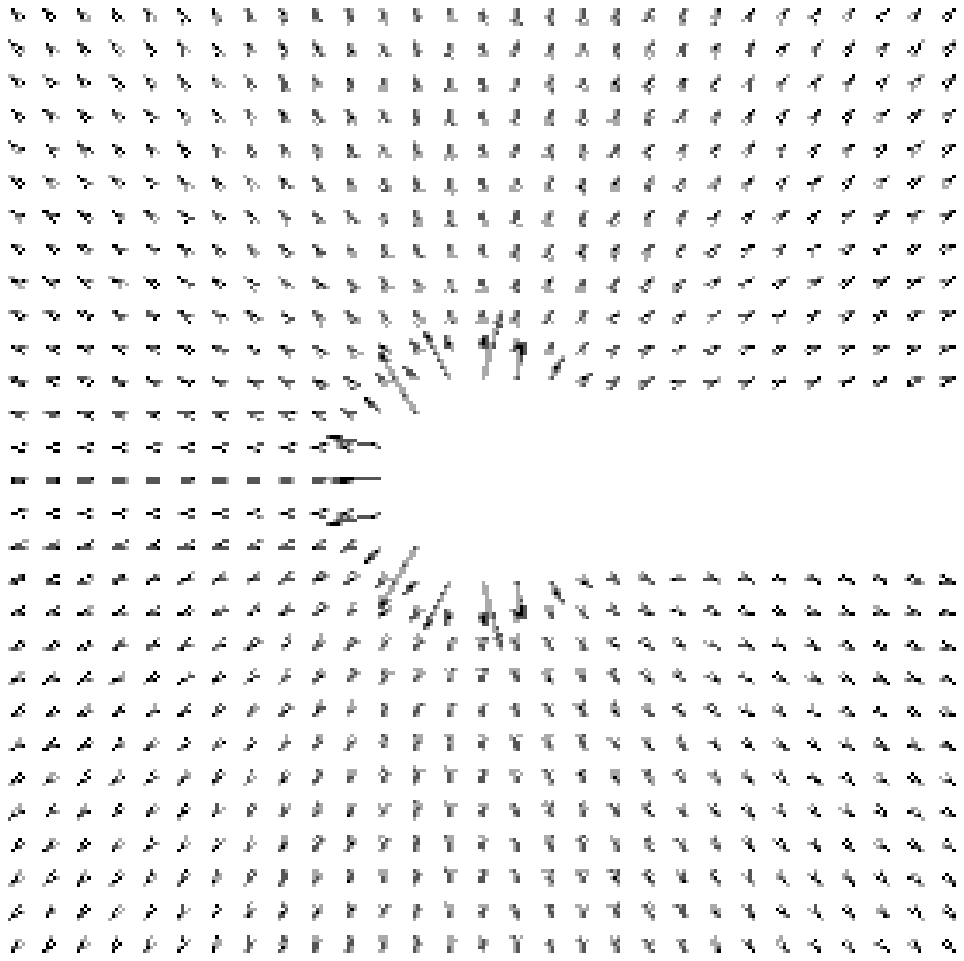}
    \label{st4k}}
  \end{minipage}
  \caption{The flow fields described by the complex velocity potential $f(z)$ and the distributions of additional term $\bm{K}$ under their flows}
  \label{stv}
\end{figure}

Fig. \ref{stv} shows the behaviors of the extra force $\bm{K}$ 
under their example flows. 
The magnitude of $\bm{K}$ depends on the size of non-commutative parameter $\theta$ and the parameters $m,\ \kappa,\ c$ in $f(z)$, but 
the shape of the distribution is the same.
\\
\section{Method of the simulation} 
In calculating the equation of motion
\begin{eqnarray*}
&&\rho\frac{D \bm{v}}{Dt}+\nabla p -\eta \Delta \bm{v}=\bm{K}\\
&&K_i=\rho\ \frac{\theta^2}{24}
\left( \frac{\partial^3 v_i }{\partial x^3 }\frac{\partial^2 v_x  }{\partial y^2}
+3\frac{\partial^3 v_i}{\partial^2 x \partial y} \frac{\partial^2 v_y }{\partial y^2}
+3\frac{\partial^3 v_i}{\partial x\partial y^2}\frac{\partial^2 v_x }{\partial x^2}
+\frac{\partial^3 v_i }{\partial y^3}\frac{\partial^2 v_y }{\partial x^2}
 \right),
\end{eqnarray*}
we use the fractional step method \cite{Smith} \cite{Kuwahara_Kawamura} , one of the standard numerical methods to solve incompressible Navier-Stokes equation. Namely, to obtain the data at (n+1)-th time step from the data at (n)-th step, 
\begin{eqnarray}
v_i^{(n+1)}&=&v_i^{*}-\frac{\Delta t}{\rho}\ \nabla p,\label{dif}\\
v_i^{*}&=&v_i^{(n)} + \frac{\Delta t}{\rho}
\left[\ -(\bm{v}\cdot \nabla) v_i + \eta \Delta v_i+K_i\ \right]^{(n)}.\nonumber
\end{eqnarray}
Applying the equation of continuity to (\ref{dif}), the Poisson equation
\begin{equation}
\nabla^2 p =\frac{\rho}{\Delta t}\ \nabla\cdot \bm{v}^{*}
\label{Po}
\end{equation}
is derived and the pressure $p$ is obtained by solving  (\ref{Po})  by the iteration method. 
The advective term is calculated by third-order accurate upwind difference method.
Third order differential factors in $K$ are  expressed as
\begin{eqnarray}
K_{i} &=&
\rho \ \frac{\theta^2}{24}\times
\bigg\{
\left(\frac{\partial}{\partial x}\frac{\partial^2 v_{i}}{\partial x^{2}}\right)
\frac{\partial^2 v_{x} }{\partial y^{2}}\nonumber\\
&+&3\left(\frac{\partial}{\partial y} \frac{\partial^2 v_{i}}{\partial x^{2}}\right)
 \frac{\partial^2 v_{y} }{\partial y^2} 
+3 \left(\frac{\partial}{\partial x}\frac{\partial^2 v_{i}}{ \partial y^{2}}\right)
 \frac{\partial^2 v_{x} }{\partial x^2} 
+ \left(\frac{\partial}{\partial y}\frac{\partial^2 v_{i} }{\partial y^{2}}\right)
 \frac{\partial^2 v_{y}}{\partial x^{2}}
\bigg\}. 
\end{eqnarray}
Here the choice of {\color{black}grid} points is changed depending on the distance from the boundaries.
In our calculation, the higher order terms of $\theta$ are ignored, by considering that even though $\theta^2$ terms are the lowest order ones, but are quite important to see the effect of the non-commutativity. 

The boundary conditions are taken as follows: On the wall boundary, the velocity $\bm{v}$ is set to be zero (no-slip) and {\color{black}the vertical component of gradient of the pressure $(\nabla p)_{\bot}$} is zero.  At the entrance of the flow at $x$ = 0 we fix the fluid velocity to be a special value and $(\nabla p)_{\bot}$ = 0, while at the exit of the flow at $x = DX$, we take the free outflow with $p$ = 0.

In this study, unequally spaced {\color{black}grid} is used.  See Fig. \ref{lat}. 
The horizontal line and the vertical line are parametrized by $0<x<DX$ and $0<y<DY$, respectively, with $DX$ = 64 cm and $DY$ = 21 cm, and 
total grid points
is 175 $\times$ 71. The flow is assumed to 
be uniform along the left boundary
\begin{eqnarray}
v_y|_{x=0}=0, ~\mathrm{and}~~ v_x|_{x=0}=-U_0 \times\left(y^2-y\ DY\right).
\end{eqnarray}
An obstacle is placed at 15 cm down stream, and so the flow is separated by this obstacle and is guided to the upper or the lower open channels. We call these open channels as slits. The upper slit is small (S), while the lower slit is large (L). The aperture of the small slit (S) is $A_S$ = 2 cm and that of the large slit (L) is $A_L$ = 8 cm. The vertical length of the obstacle is $A$ = 11 cm. Both ends of the obstacle are semicirculars with the radius of 2 cm. The minimum size of the {\color{black}grid} spacing is 0.167 cm. The step size of time is $10^{-3}$ [s] and it carries on 2000000 steps to calculate 2000 [s] flow.  

We have considered the flow 
analysis given above, with the following intention: The hydrodynamics on the non-commutative space may give the minimum area $\theta$ to the size of the constituent particles, so that  if $\theta$ is introduced, then the flow rate passing through the small slit (S) is reduced and that passing through the large slit (L) is increased while keeping the total flow rates constant.

{\begin{figure}[h]
\centering
\includegraphics[width=140mm]{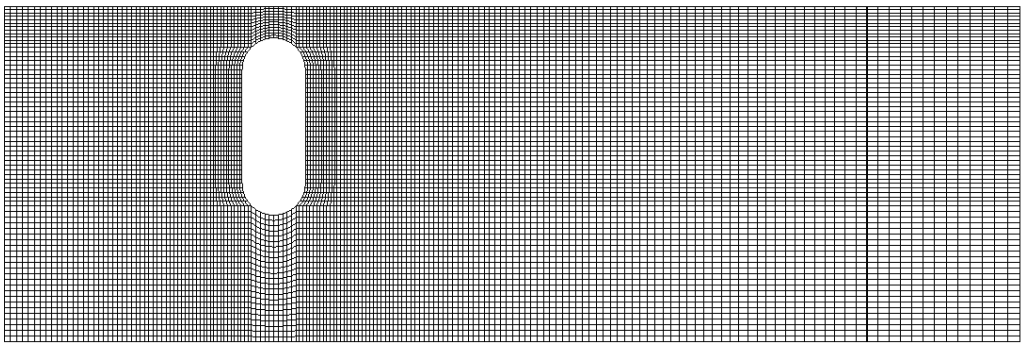}
\caption{The {\color{black}grid} used for simulation.}
\label{lat}
\end{figure}}

We fix the Reynolds number Re = 700. It is because in the following analysis, 
the activity of the vortices is important to see the non-commutativity effects.
So, we wish to take a large value of Re, but the larger Re $>$ 2000 introduces the turbulent flow, and hence we fix it to 700. 
If we consider the water flow, then $\rho=10^3 \mathrm{kg}/\mathrm{m}^3$ and $\eta=8.85 \times 10^{-4} \mathrm{Pa}\cdot \mathrm{s}$.  In this case, if $U_{\star}$ is chosen to $U_0=$ 0.55 cm/s and $L_{\star}$ is chosen to the obstacle length $A=$11 cm.

\newpage
\section{Results of the simulation}
\begin{figure}[!h]
\centering
\includegraphics[width=100mm]{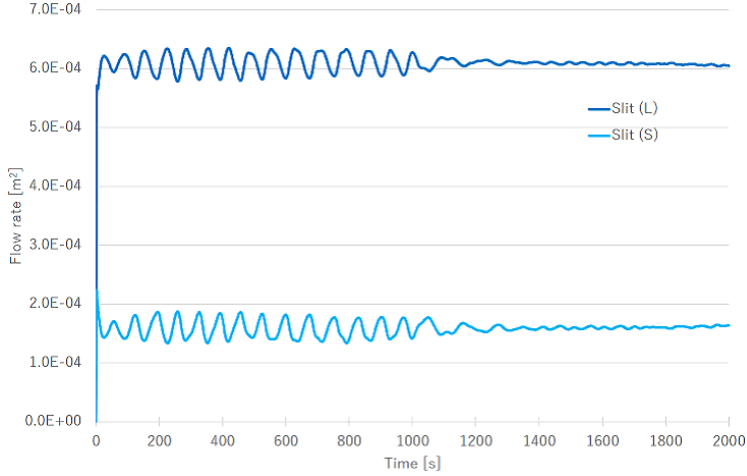}
\caption{Flow rates at the slit (S) (light blue) and (L) (dark blue) in $\sqrt{\theta}=0$ cm case.}
\label{r0}
\end{figure}

\begin{figure}[!h]
\centering
\includegraphics[width=120mm]{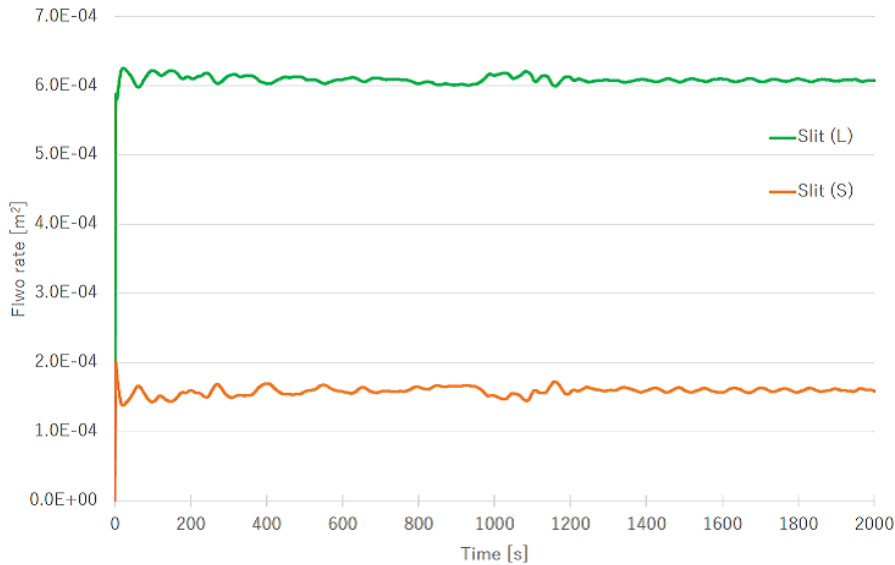}
\caption{Flow rates at the slit (S) and (L) in $\sqrt{\theta}=0.29$ cm case.}
\label{r2.9}
\end{figure}

In order to compare the flowability through the slits between the usual flow and the non-commutative flow, the flow rates at slit (S) and (L) are estimated without and with $\theta$ in Fig. \ref{r0} and Fig. \ref{r2.9}, respectively.

First look at Fig. \ref{r0}, $\sqrt{\theta}=0$ cm which is the usual flow.  There exists the fluctuation of flow rates in the beginning, being regularly and periodically depicted like a sine curve, but afterwards for $t > 1100$ [s], the oscillation is damped.  This is a characteristic of the usual flow.
The following figures show the snapshots of the flow at $t=$495 [s], 528[s], 1025 [s] and 2000 [s].

\begin{figure}[!h]
  \setlength{\subfigwidth}{.5\linewidth}
  \addtolength{\subfigwidth}{.5\subfigcolsep}
  \begin{minipage}[b]{\subfigwidth}
    \centering\subfigure[Snapshot at $t=$ 495 {[s]} in the case of $\theta$=0 cm]{\includegraphics[width=7.0cm]{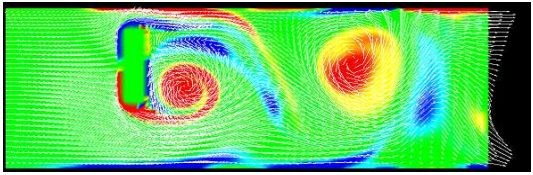}
   \label{r0a}}
  \end{minipage}
  \begin{minipage}[b]{\subfigwidth}
    \centering\subfigure[Snapshot at $t=$528 {[s]} in the case of $\theta$=0 cm]{\includegraphics[width=7.0cm]{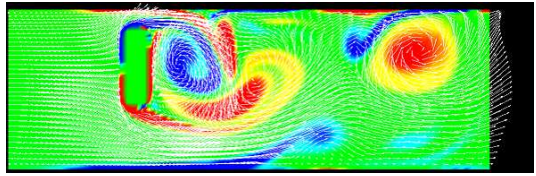}
   \label{r0b}}
  \end{minipage}

  \begin{minipage}[b]{\subfigwidth}
    \centering\subfigure[Snapshot at $t=$1025 {[s]} in the case of $\theta$=0 cm]{\includegraphics[width=7.0cm]{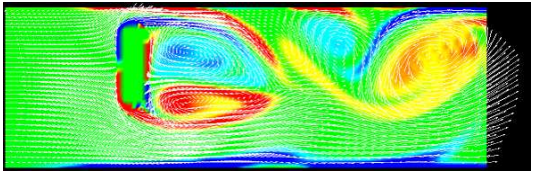}
   \label{r0c}}
  \end{minipage}
  \begin{minipage}[b]{\subfigwidth}
    \centering\subfigure[Snapshot at $t=$2000 {[s]} in the case of $\theta$=0 cm]{\includegraphics[width=7.0cm]{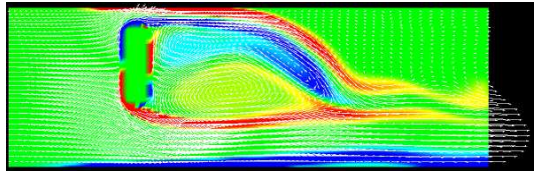}
   \label{r0d}}
  \end{minipage}

  \caption{The velocity field and the vorticity of the $\sqrt{\theta}$ = 0 cm flow.  The velocity at each points is depicted by the white arrow while the vorticity is represented by the color gradation, that is, blue is negative vorticity (counter-clockwise rotation), red is positive vorticity (clockwise rotation) and green is the medium.}
  \label{r0v}
\end{figure}

 The velocity field and the vorticity of $\sqrt{\theta}$ = 0 cm flow is shown  in Fig. \ref{r0v}, where the velocity at each points is depicted by the white arrow while the vorticity is represented by the color gradation, that is, blue is negative vorticity (anti-clockwise rotation), red is positive vorticity (clockwise rotation) and green is the medium. While the flow rates at slit (S) and (L) oscillate, vortices are periodically generated near the slits and the other boundaries, and flow out into the down stream as shown in Fig. \ref{r0a}.

\newpage
In the periodical oscillation phases of (a): 495 [s] and (b): 528 [s], there appear a number of red regions and a number of blue regions which means the vortices are periodically produced and flow out. The spiral just behind the obstacle is changed from red at (a) to blue at (b). 

However at (c): 1025 [s], the blue regions and red regions begin individually connected to form the bands, and  the so called ``two attached eddies" appear behind the obstacle. Finally at (d): 2000 [s], the flow becomes stable.  This shows the damping of the oscillation of the flow rates.
\newpage
Now we will show the flow with $\sqrt{\theta}=0.29$ cm.
Fig. \ref{r2.9} which gives the flow rates at slit (S) (brown) and slit (L) (green).

This figure shows the behavior of the flow rates at slit (S) and (L) in case of $\sqrt{\theta}$=0.29 cm, or of the non-commutative flow. 
The oscillation of flow rates is observed, 
but the activity of the oscillation is weak and irregular compared to the case of $\theta=0$, vortices are observed.  The flow rates becomes stable after $t=$1300 [s].

\begin{figure}[!h]
  \setlength{\subfigwidth}{.5\linewidth}
  \addtolength{\subfigwidth}{.5\subfigcolsep}
  \begin{minipage}[b]{\subfigwidth}
    \centering\subfigure[Snapshot at $t=$ 495 {[s]} for $\sqrt{\theta}=0.29$ cm]{\includegraphics[width=7.0cm]{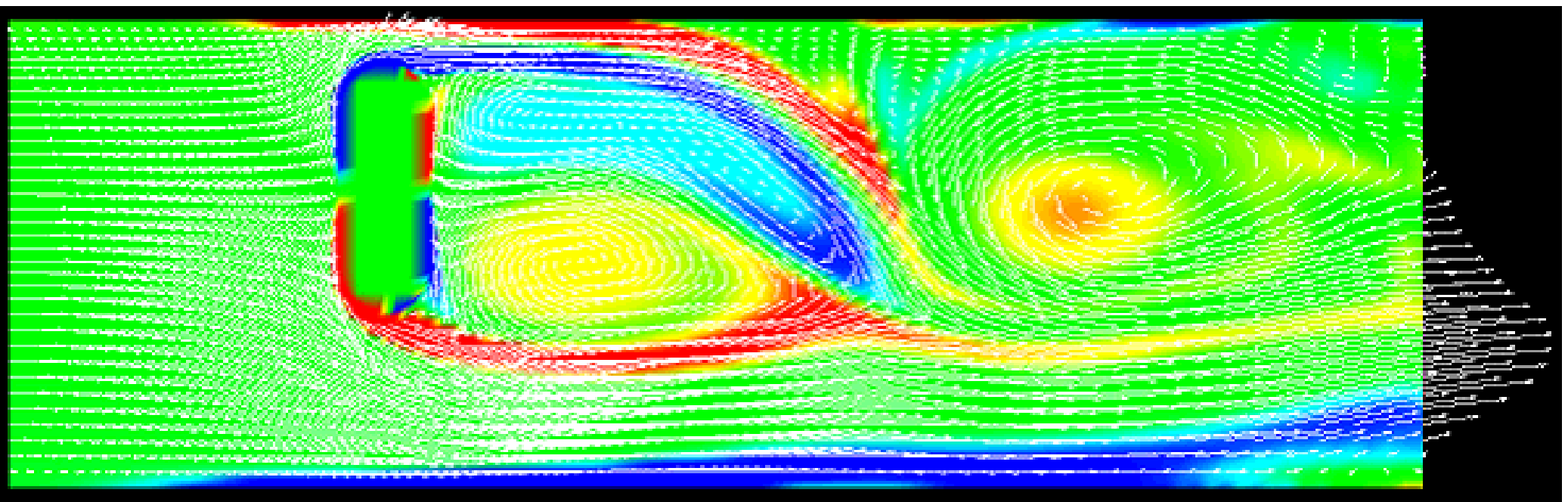}
   \label{r29a}}
  \end{minipage}
  \begin{minipage}[b]{\subfigwidth}
    \centering\subfigure[Snapshot at $t=$850 {[s]} for $\sqrt{\theta}=0.29$ cm]{\includegraphics[width=7.0cm]{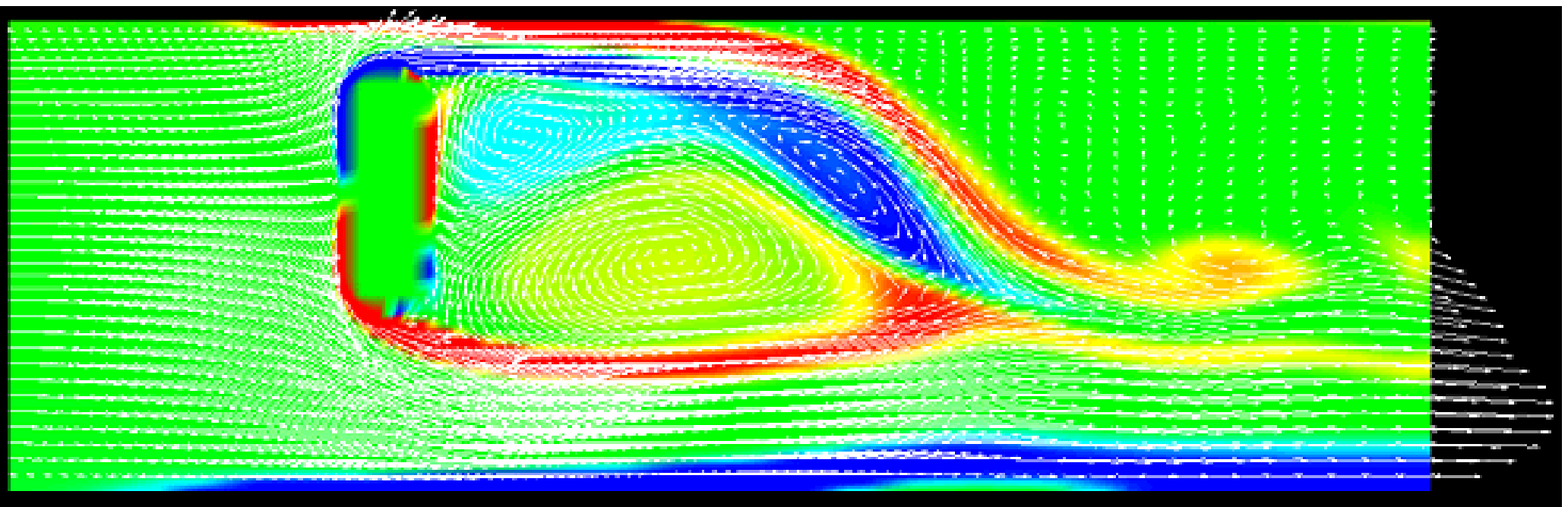}
   \label{r29b}}
  \end{minipage}

  \begin{minipage}[b]{\subfigwidth}
    \centering\subfigure[Snapshot at $t=$1028 {[s]} for $\sqrt{\theta}=0.29$ cm]{\includegraphics[width=7.0cm]{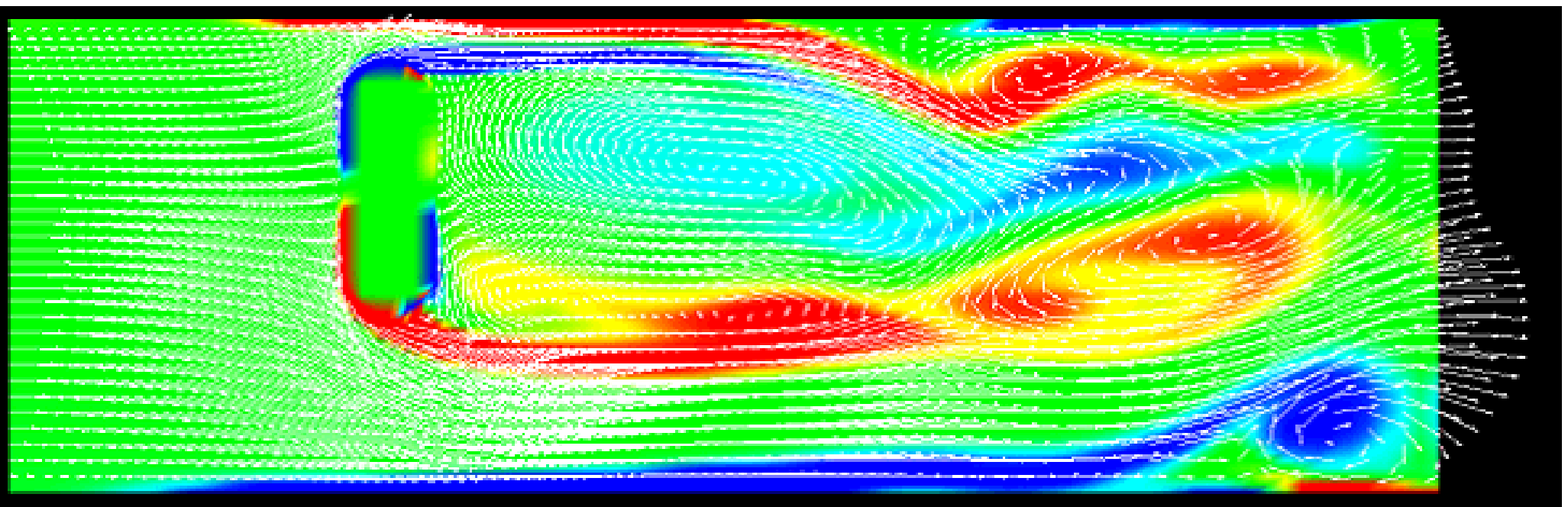}
   \label{r29c}}
  \end{minipage}
  \begin{minipage}[b]{\subfigwidth}
    \centering\subfigure[Snapshot at $t=$ 2000 {[s]} for $\sqrt{\theta}=0.29$ cm]{\includegraphics[width=7.0cm]{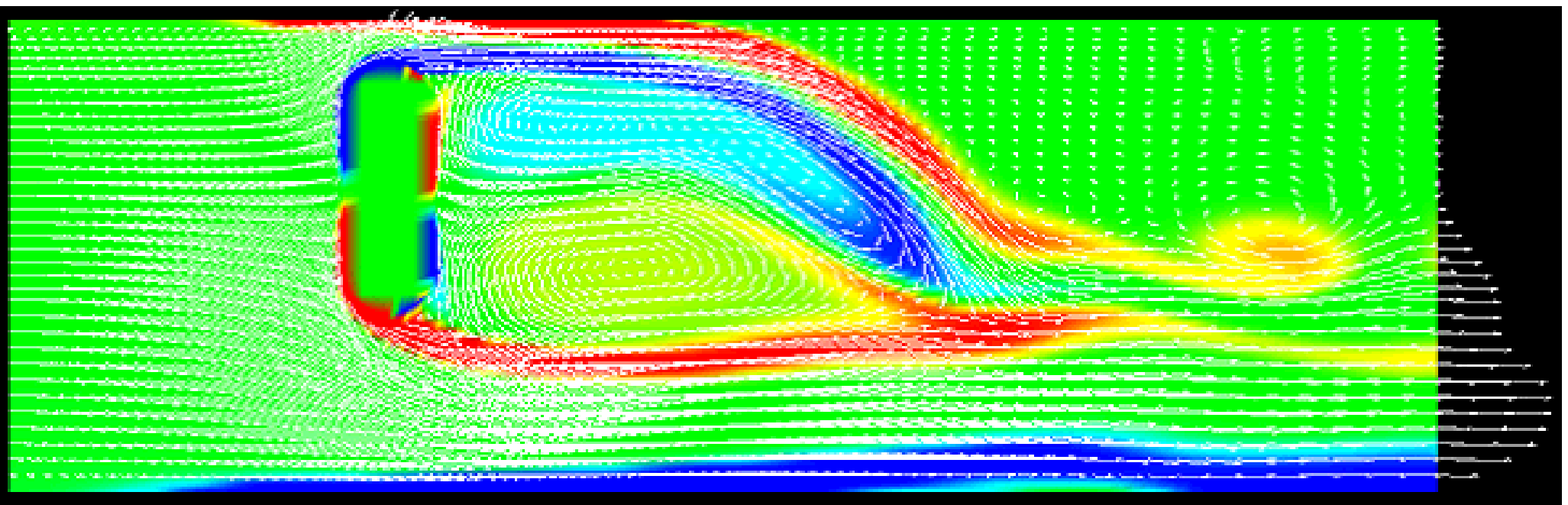}
   \label{r29d}}
  \end{minipage}

  \caption{Velocity and vortices for the non-commutative flow with $\sqrt{\theta}=0.29$ cm.}
  \label{r29v}
\end{figure}

\begin{figure}[h]
\centering
\includegraphics[width=12cm]{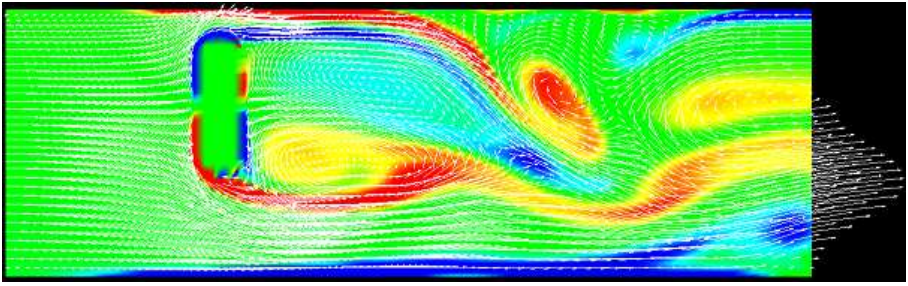}
\caption{Snapshot at $t$ = 1143 [s] in the case of $\sqrt{\theta}=0.29$ cm}
\label{r29v2}
\end{figure}

This behavior can be understood from the snapshot Figs. \ref{r29v}
taken at $t=$ 495, 850, 1028 and 2000 [s].  The figures show that the pattern of the vortices are almost stable from the beginning.  There is no active production of vortices just after the obstacle as was observed in the normal flow.  The extra force disturbs the active production of vortices behind the obstacle.  The oscillation arises from the activity of vortices in the down stream.  At (c):1028 [s] 
{\color{black}the vortices in the down stream collapse}, but after 1300 [s], the whole stream becomes stable.
At around $t=$ 1143 [s], 
{\color{black}the flowing vortices bocome staying behind of the obstacle again}, where the activity of the vortex can be seen in the snapshot in Fig. \ref{r29v2} at this moment. 
\newpage

\begin{figure}[h]
\centering
\includegraphics[width=120mm]{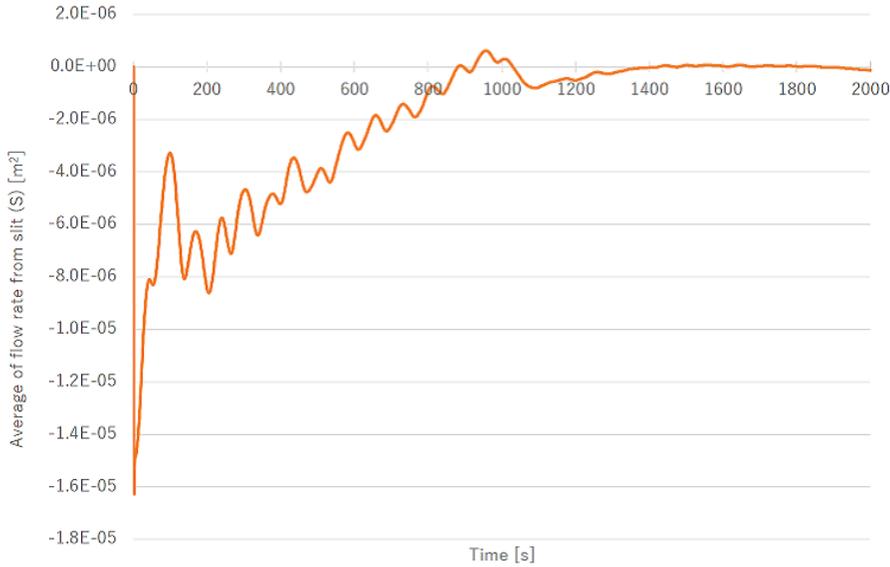}
\caption{Average of flow rate from slit (S) at that time.}
\label{ave}
\end{figure}

The averaged difference of the flow rates $\langle \Delta S \rangle$ at the small slit with and without $\theta$, that is, $\langle \Delta S \rangle=\langle S(\sqrt{\theta}=0.29\ \mathrm{cm}) - S(\sqrt{\theta}=0\ \mathrm{cm}) \rangle $ is depicted in Fig. \ref{ave}, where the flow rates are averaged between 0 [s] and $t$ [s].
Before $t$ = 1000 [s], the flow rate at slit (S) becomes smaller for $\sqrt{\theta}=0.29$ cm than that for $\theta = 0$, which matches to the intuitive understudy of the size of the fluid particle.
However, the discrepancy disappears 
after $t$ = 1000 [s].
\newpage


\section{Conclusion and Discussion}
In this paper a two-dimensional simulation is performed for the hydrodynamics on the non-commutative space.  The parameter characterizing the non-commutativity of the space, $\theta$, gives the uncertainty relation of the space, or the minimum size of the fluid particle $\Delta x \Delta y > \theta/2$.

We have prepared the flow path of 21 cm wide and 64 cm length where the obstacle is placed in the middle of the flow, and so the flow should passes through the small 2 cm slit (S) or the large 8 cm slit (L).  The simulation is done for the water with Reynolds number Re = 700, where the maximal incoming velocity is 0.55 cm/s.
Re = 700 is so chosen as the activity of vortex is large, but is not too large to introduce the turbulent flow.

The results we obtained are as follows:

(1) For the usual flow without $\theta$, the flow rate at the small slit (S) oscillates like a sine curve, but the amplitude of the oscillation diminishes slowly after $t=$1200 [s].

(2) For the non-commutative flow with $\sqrt{\theta}=$0.29 cm, the oscillation of the flow rate is small and irregular.  The fluctuation is stabilized after $t=$1200 [s]. 

(3) The pattern of the flow is correlated to the activity of the vortices. 
In the usual flow the oscillation of the flow late is diminished after the certain time has passed, and this change is related to the occurrence of the ``two attached eddies".  In the non-commutative flow this ``two attached eddies" appears from the beginning, and hence the flow rates do not actively oscillate.  So, the difference between the usual flow and the non-commutative flow comes from the difference of activities of vortices in the down stream.

As is understood from the analysis in the perfect fluid, the extra force $\bm{K}$, relevant to the non-commutativity, has the tendency to interrupts the flowing out into the sink, and to expand the size of the vortex in the radial direction.  These effects are small for a measure of the particle size $/$ the size of the small slit, Sm $=\sqrt{\theta}/A_S\ =$ 0.29 cm $/$ 2 cm 
$\simeq$ 
 0.15, but they become extremely large near the singular positions such as the position of sink and source and the center of the vortex.
Therefore, it is important to know how the vortices are created from the boundary layers, especially from the boundary layer of the obstacle near the slits in both cases with and without $\theta$, as well as to know the difference of the interaction of vortices both in the usual flow and in the non-commutative flow.  For this we have to know the vorticity transportation equation with the extra force $\bm{K}$, being proportional to the non-commutative parameter squared, $\theta^2$. 

This can be done easily. Using the complex notations, the Navier-Stokes equation with $\bm{K}$ reads
\begin{eqnarray}
&&\rho \left[ 2i \frac{\partial}{\partial t}(\partial \varphi) + 4 \partial \varphi \left(\overleftarrow{\partial} \overrightarrow{\bar{\partial}} - \overleftarrow{\bar{\partial}} \overrightarrow{\partial} \right) \varphi \right] \nonumber \\ 
&&= -2 \partial p + 8i \eta (\partial \bar{\partial})(\partial \varphi) - \rho \frac{2}{3} \theta^2  \partial \varphi \left(\overleftarrow{\partial} \overrightarrow{\bar{\partial}} - \overleftarrow{\bar{\partial}} \overrightarrow{\partial} \right)^3 \varphi.
\end{eqnarray}
Taking the rotation of this equation, $rot A = \bar{\partial} A- \partial \bar{A}$,  we have the following equation
\begin{eqnarray}
&&\left(\rho \frac{\partial}{\partial t} - \eta \Delta  \right) \omega=\rho \frac{\partial \omega}{\partial t} - 4 \eta (\partial \bar{\partial}) \omega \nonumber \\
&&= 2 \rho \omega i \left(\overleftarrow{\partial} \overrightarrow{\bar{\partial}} - \overleftarrow{\bar{\partial}} \overrightarrow{\partial} \right) \varphi + \frac{\rho}{3} \theta^2 \omega i \left(\overleftarrow{\partial} \overrightarrow{\bar{\partial}} - \overleftarrow{\bar{\partial}} \overrightarrow{\partial} \right)^3 \varphi,
\end{eqnarray}
where $\omega$ is the vorticity defined by 
\begin{eqnarray}
\omega=\partial_x v_y-\partial_y v_x=-4 (\partial \bar{\partial})\varphi=-\Delta \varphi. \label{def of vorticity}
\end{eqnarray}
If $\theta=0$, this is the usual vorticity transportation equation which describes the disscipation by the viscosity $\eta$.

On the non-commutative space, the vorticity transpiration equation is modified so as to include the $\theta^2$ term.

The stream function can be expressed in terms of the vorticity by Eq. (\ref{def of vorticity}) as
\begin{eqnarray}
\varphi(z, \bar{z})=\int d^2u \ln \left( (z-u)(\bar{z}-\bar{u}) \right) \omega(u, \bar{u}).
\end{eqnarray}

Then, the vorticity transportation equation can be written on the non-commutative space as follows:
\begin{eqnarray}
&&\left(\rho \frac{\partial}{\partial t} - \eta \Delta  \right) \omega \nonumber \\
&=&2 \rho \int d^2 u ~\left\{ i \left(\partial_z \bar{\partial}_u - \bar{\partial}_z \partial_u \right) + \frac{\theta^2}{3!} i \left(\partial_z \bar{\partial}_u- \bar{\partial}_z \partial_u \right)^3 \right\}\nonumber \\
&&~~~~~~~~~~~~~~\times \left\{\omega(z, \bar{z}) \ln \left( (z-u)(\bar{z}-\bar{u}) \right)\omega(u, \bar{u})\right\}.
\end{eqnarray}

The interaction between vortices is modified naturally in the hydrodynamics on the non-commutative space.  

The remained problem to be studied is by using the creation rate and the interaction of vortices to clarify the simulation results obtained in this paper.

A preliminary stage of this paper was described in the thesis by one of the authors (M.S.) \cite{D}.

%
%
\section*{Acknowledgements}
The authors are grateful to Kazuharu Bamba, Gi-Chol Cho, Takeo Inami, Katsuya Ishii, Satoshi Iso, Takashi Koide, Satoko Komurasaki, Shin Nakamura, and Ko Okumura for fruitful discussions.
The flow fields were visualized by "Clef3D" made by Institute of Computational Fluid Dynamics.
%
%


\newpage
\begin{thebibliography}{99}
\bibitem{Nambu dynamics} Y. Nambu, Phys.\ Rev.\  {\bf D7}, 2405 (1973).
\bibitem{Nambu 1} Y. Nambu: a talk at International Workshop: Extra Dimensions in the Era of the LHC, Dec. 12-14 (2011);\\
Y. Nambu: a talk at International Symposium on Research Frontiers on Physics, Earth and Space Science, Dec. 17-18 (1913).  
\bibitem{Saitou 1} M. Saitou, K. Bamba and A. Sugamoto: PTEP 103B03 (2014).
\bibitem{Moyal} J.~E.~Moyal, ``Quantum Mechanics as a Statistical Theory", in the Proceedings of Cambridge Philosophical Society, Vol {\bf 45}, 99 (1949).
\bibitem{Smith} G. D. Smith: ``Numerical solution of partial differential equations" 
(Oxford University Press, 1985)
\bibitem{Kuwahara_Kawamura} T. Kawamura and K. Kuwahara: ``Computation of high Reynolds number flow around circular cylinder with surface roughness", AIAA Paper 84-0340 (1984);\\
K. Kuwahara and T. Kawamura: ``Fluid computation and difference calculus" (in Japanese) (Asakura Publishing Co., 2005).
\bibitem{D} M. Saitou, ``Hydrodynamics on non-commutative space" (in Japanese), \\
(March. 2016) http://hdl.handle.net/10083/59556
\end{thebibliography}
\end{document}